\documentclass[slac_one]{revtex4}
\usepackage{graphicx}
\usepackage{fancyhdr}
\begin{document}

\title{Measurement of Charged Current Charged Single Pion Production in SciBooNE}

%

\author{K.~Hiraide for the SciBooNE collaboration}
\affiliation{Department of Physics, Kyoto University, Kyoto 606-8502, Japan}

\begin{abstract}
The SciBooNE experiment is designed to measure neutrino cross sections on carbon around
one GeV region. Charged current single charged pion production is a dominant background process
for $\nu_\mu \to \nu_x$ oscillation experiments with a few-GeV neutrino beam, and thus a precision measurement of the cross section is essential. This article reports preliminary results
on this process from SciBooNE.
\end{abstract}

\maketitle

\thispagestyle{fancy}


\section{INTRODUCTION} 
Neutrino-induced charged current single charged pion production (CC-1$\pi^+$) is dominated
by baryonic resonance excitation off a single nucleon bound in a nucleus in the neutrino
energy region of a few GeV. The resonance state is followed by its prompt decay into a nucleon
and a pion in the final state. The process is written as $\nu_\mu N \to \mu^- N \pi^+$,
where $N$ is proton or neutron.
In addition to this reaction, neutrinos can produce pions by interacting coherently
with the nucleons forming the target nucleus. 
The process is expressed as $\nu_\mu A \to \mu^- A \pi^+$, where $A$ is a nucleus.

Understanding the cross sections of these processes is important to study $\nu_\mu \to \nu_x$
oscillation near one GeV. In such oscillation experiments, a distortion in the
$\nu_\mu$ energy spectrum is measured with charged current quasi-elastic (CC-QE) interactions,
$\nu_\mu n \to \mu^- p$, by reconstructing neutrino energy from the measured muon momentum and angle. 
The background to this channel is dominated by CC-1$\pi^+$ events in which the pion is not observed
so that the final state looks like a CC-QE interaction. This mis-identification comes from 
the lack of the final state $\pi^+$ detection due to low energy as well as pion absorption
inside the nucleus. In case of the T2K experiment~\cite{Itow:2001ee}, a next generation long baseline
neutrino oscillation experiment, the CC-1$\pi^+$/CC-QE cross section ratio is required to be understood
at 5\% level to keep the resulting error on the oscillation parameters comparable to that due to
statistical uncertainties~\cite{AguilarArevalo:2006se}.
However, the current knowledge of the cross section ratio has been limited by low statistics
and large systematic uncertainty, and therefore a precision measurement of the cross section
is essential. 
 
In addition, recent results on coherent pion production have induced interest of the neutrino
physics community. The non-existence of CC coherent pion production in a 1.3~GeV wide-band neutrino
beam has been reported by K2K~\cite{Hasegawa:2005td}, while there exist CC coherent pion production
positive results at higher neutrino energies. On the one hand, evidence for NC coherent pion production
in the similar neutrino energy has been recently reported by MiniBooNE~\cite{AguilarArevalo:2008xs}.

\section{SCIBOONE EXPERIMENT}
The SciBooNE experiment~\cite{AguilarArevalo:2006se} is designed to measure the neutrino
cross sections on carbon in the one GeV region. The experiment uses the Booster Neutrino Beam
(BNB) at Fermilab. The primary proton beam, with kinetic energy 8~GeV, is extracted to strike
a 71~cm long, 1~cm diameter beryllium target. Each beam spill consists of 81 bunches of protons,
containing typically $4\times 10^{12}$ protons in a total spill duration of 1.6~$\mu$sec.
The target sits at the upstream end of a magnetic focusing horn that is pulsed with
approximately 170~kA to focus the mesons, primarily $\pi^+$, produced by the $p\rm{-Be}$ interactions.
In a 50 m long decay pipe following the horn, $\pi^+$ decay and produce neutrinos, before
the mesons encounter an absorber. The flux is dominated by muon neutrinos (93\% of total),
with small contributions from muon antineutrinos (6.4\%), and electron neutrinos and
antineutrinos (0.6\% in total). The flux-averaged mean neutrino energy is 0.7 GeV.
When the horn polarity is reversed, $\pi^-$ are focused and hence a predominantly antineutrino
beam is created.

The SciBooNE detector is located 100 m downstream from the neutrino production target.
The detector complex consists of three sub-detectors: a fully active fine grained
scintillator tracking detector (SciBar), an electromagnetic calorimeter (EC) and
a muon range detector (MRD). The SciBar detector consists of 14,336 extruded plastic
scintillator strips, each $1.3\times 2.5\times 300$~cm$^3$. The scintillators are arranged
vertically and horizontally to construct a  $3\times 3\times 1.7$ m$^3$ volume with a total
mass of 15 tons. Each strip is read out by a wavelength-shifting (WLS) fiber attached
to a 64-channel multi-anode PMT (MA-PMT). Charge and timing information from each MA-PMT
is recorded by custom electronics. The minimum length of a reconstructed track is 8~cm
which corresponds to a proton with momentum of 450~MeV/$c$.
The EC is installed downstream of SciBar, and consists of 32 vertical and 32 horizontal
modules made of scintillating fibers embedded in lead foils. Each module has dimensions
of $4.0\times 8.2\times 262$ cm$^3$, and is read out by two 1'' PMTs on both ends.
The EC has a thickness of $11X_0$ along the beam direction to measure $\pi^0$ emitted
from neutrino interactions and the intrinsic $\nu_e$ contamination.
The energy resolution is $14\%/\sqrt{E\rm{[GeV]}}$.
The MRD is located downstream of the EC in order to measure the momentum of muons up to
1.2 GeV$/c$ with range. It consists of 12 layers of 2''-thick iron plates sandwiched
between layers of 6 mm-thick plastic scintillator planes. The cross sectional area of
each plate is $305\times 274$ cm$^2$. The horizontal and vertical scintillator planes
are arranged alternately, and the total number of scintillators is 362.

The experiment took both neutrino and antineutrino data from June~2007 until August~2008.
In total, $2.64\times 10^{20}$~POT were delivered to the beryllium target during the SciBooNE data
run. After beam and detector quality cuts, $2.52\times 10^{20}$~POT are usable for physics
analyses; $0.99\times 10^{20}$~POT for neutrino data and $1.53\times 10^{20}$~POT for antineutrino
data. Preliminary results from the full neutrino data sample are presented in this paper.

\section{CC-1$\pi^+$ EVENT SAMPLE}
The experimental signature of CC single charged pion production is the existence of two and only
two tracks originating from a common vertex, both consistent with minimum ionizing particles
(a muon and a charged pion). Even in case of a CC resonant pion event with proton in the final
state, $\nu p\to \mu^- p\pi^+$, the proton is often not reconstructed due to its low energy.

To identify CC events, we search for tracks in SciBar matching with a track or hits in the MRD.
Such a track is defined as a SciBar-MRD matched track. The most energetic SciBar-MRD matched
track in any event is considered as the muon candidate. The matching criteria impose a muon momentum
threshold of 350~MeV/$c$. The neutrino interaction vertex is reconstructed
as the upstream edge of the muon candidate. We select events whose vertices are in the SciBar fiducial
volume (FV), $2.6 \ {\rm m}\times 2.6 \ {\rm m}\times 1.55 \ {\rm m}$, a total mass of 10.6~tons.
Finally, event timing is required to be within 2~$\mu$sec beam timing window.
The cosmic-ray background contamination in the beam timing window is only 0.5\%, estimated using
a beam-off timing window. Approximately 30,000 events are selected as our standard CC sample.
According to the MC simulation, the selection efficiency and purity of true $\nu_\mu$ CC events
are 28\% and 93\%, respectively.

In order to reconstruct muon momentum from its range, the muon candidate is required to stop in the MRD.
Once the muon candidate and the neutrino interaction vertex are reconstructed, we search for other
tracks originating from the vertex. Events with two and only two vertex-matched tracks are selected.
The SciBar detector has the capability to distinguish protons from muons and pions using $dE/dx$.
The particle identification variable, Muon Confidence Level (MuCL) is related to the probability
that a particle is a minimum ionizing particle (MIP) based on the energy deposition.
The confidence level at each plane is first defined as the fraction of events in the expected
$dE/dx$ distribution of muons above the observed value. The expected $dE/dx$ distribution of
muons is obtained by using cosmic-ray muons. Each plane's confidence level is combined to form
a total confidence level, assuming the confidence level at each layer is independent. 
The particle identification is applied to both tracks to select events with two MIP-like tracks
($\mu+\pi$). The probability of mis-identification is estimated to be 1.1\% for muons and 12\%
for protons.

After selecting $\mu+\pi$ events, the sample still contains CC-QE events in which a proton is
mis-identified as a MIP-like track. We reduce CC-QE background by using kinematic information.
Since the CC-QE interaction is a two-body interaction, one can predict the proton
direction from the measured muon momentum and muon angle. We define an angle called $\Delta\theta_p$
as the angle between the expected proton track direction and the observed second track direction.
In CC-QE events, the angle $\Delta \theta_p$ is expected to be small, and therefore we select events
in which $\Delta \theta_p$ is greater than 20~degrees to reject CC-QE background.
With this selection, 48\% of CC-QE events in the $\mu+\pi$ sample are rejected.

\begin{figure}[tbp]
  \begin{center}
    \includegraphics[keepaspectratio=true,height=70mm]{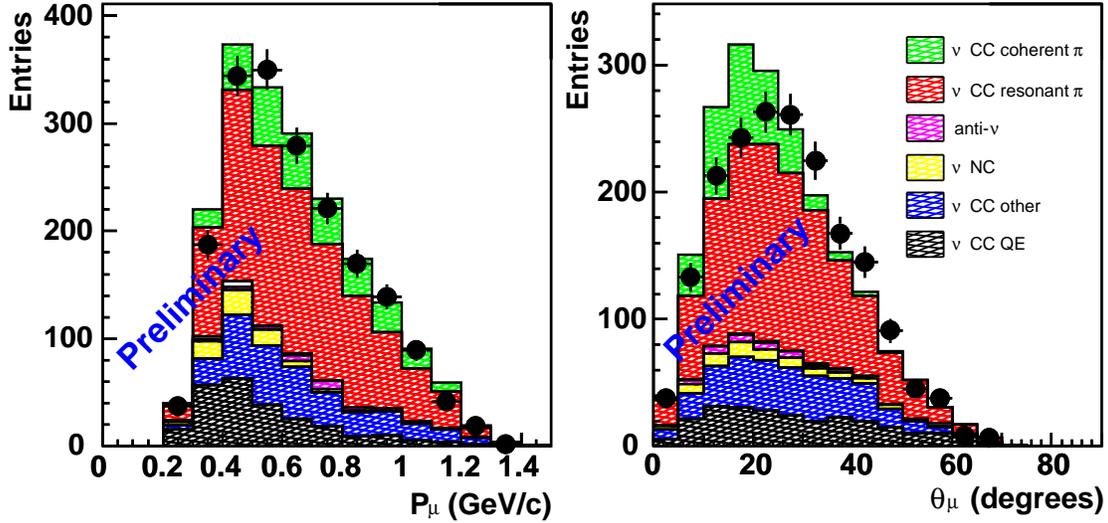}
  \end{center}
  \caption{(Color online) Reconstructed muon momentum (left) and angle with respect to the neutrino
  beam direction (right) for CC-1$\pi^+$ sample.}
  \label{fig:pmuthetamu_cc1pi.eps}
\end{figure}

We select approximately 2,000 CC-1$\pi^+$ candidates.
Figure~\ref{fig:pmuthetamu_cc1pi.eps} shows reconstructed muon momentum and angle with respect to
the neutrino beam direction for CC-1$\pi^+$ sample. For the MC simulation, the contributions from each
interaction mode are shown separately. According to the MC simulation, the fraction of CC resonant pion
production ( $\nu_\mu p\to \mu^- p\pi^+$ and $\nu_\mu n\to \mu^- n\pi^+$) and CC coherent pion production
in the sample is 34\%, 11\%, and 15\%, respectively. As seen in the figure, the deficit of data in the small
muon angle region is found. Further investigation has been performed by extracting CC coherent pion events.

\section{EXTRACTION OF CC COHERENT PION EVENTS}
CC coherent pion events are separated from CC resonant pion events by using their characteristic
kinematic information. 
Because of the small momentum transfer to the target nucleus, both the muon and pion go in the
forward direction. Events in which a pion candidate goes forward are selected.
The additional protons with momentum below the tracking threshold are detected by their large
energy deposition around the vertex, so-called vertex activity. We search for the maximum
deposited energy in a strip around the vertex, an area of $12.5 \ {\rm cm} \times 12.5 \ {\rm cm}$
in both views. Events with the energy deposition less than 10 MeV are considered to have
no vertex activity and selected.

\begin{figure}[tbp]
  \begin{center}
    \includegraphics[keepaspectratio=true,height=70mm]{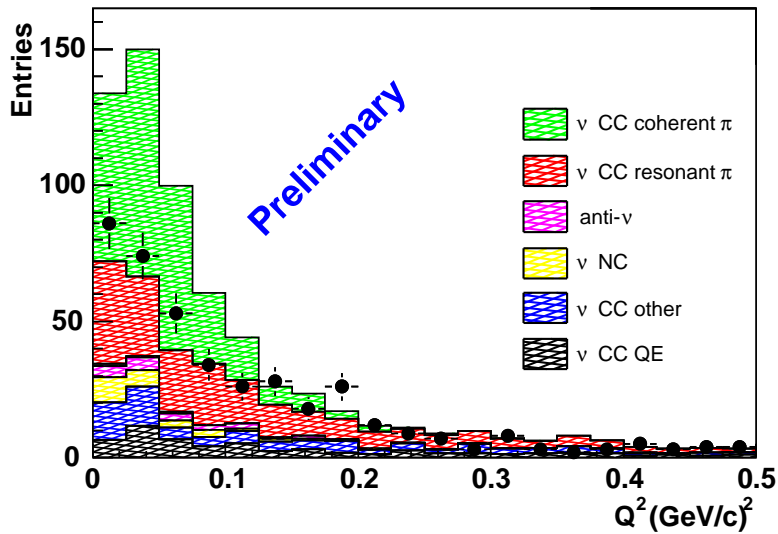}
  \end{center}
  \caption{(Color online) Reconstructed $Q^2$ assuming CC-QE kinematics for CC coherent pion sample.}
  \label{fig:q2rec_coherent.eps}
\end{figure}

Figure~\ref{fig:q2rec_coherent.eps} shows reconstructed $Q^2$ assuming CC-QE kinematics for the CC
coherent pion sample. Although a CC-QE interaction is assumed, the $Q^2$ of CC coherent pion events
is reconstructed with a resolution of 0.02~(GeV/$c$)$^2$ and a shift of -0.02~(GeV/$c$)$^2$
according to the MC simulation. Finally, events with reconstructed $Q^2$ less than 0.1~(GeV/$c$)$^2$
are selected as CC coherent pion candidates. The selection efficiency of CC coherent pion production
is 11\%, and the mean neutrino energy after accounting for the selection efficiency is 1.1~GeV,
both estimated with the MC simulation.
The observed CC coherent pion sample contains fewer events than our MC prediction which is based on
the Rein-Sehgal model with lepton mass correction~\cite{Rein:1982pf,Rein:2006di}.
Our 90\% C.L. sensitivity on the cross section ratio of CC coherent pion production to
total CC interaction is estimated to be 0.004, which corresponds to 20\% of the MC prediction.

\section{SUMMARY}
In summary, we present preliminary results on charged current single charged pion production,
analyzing the full neutrino data corresponding to $0.99\times 10^{20}$ POT.
Observed CC coherent pion sample contains fewer events than our MC prediction which is based
on the Rein and Sehgal model. Our 90\% C.L. sensitivity on the cross section ratio of CC
coherent pion production to total CC interaction is 0.004 at the mean neutrino energy of 1.1~GeV.
After measuring CC coherent pion production, the measurement on CC resonant pion production
will be performed.

\begin{acknowledgments}
The SciBooNE Collaboration gratefully acknowledge support from various grants,
contracts and fellowships from the MEXT (Japan), the INFN (Italy),
the Ministry of Education and Science and CSIC (Spain), the STFC (UK),
and the DOE and NSF (USA). The author is grateful to the Japan Society
for the Promotion of Science for support.
\end{acknowledgments}

\end{document}